# Nonlinear Transport of Graphene in the Quantum Hall Regime

Shibing Tian[+,1], Pengjie Wang[+,1], Xin Liu[1], Junbo Zhu[1], Hailong Fu[1], Takashi Taniguchi[2], Kenji Watanabe[2], Jian-Hao Chen[*,1,3], Xi Lin[*,1,3]

[1] International Center for Quantum Materials, Peking University, Beijing 100871, China

[2] High Pressure Group, National Institute for Materials Science, 1-1 Namiki, Tsukuba, Ibaraki 305-0044, Japan

[3] Collaborative Innovation Center of Quantum Matter, Beijing 100871, China

[+]These two authors contributed equally to this work
[*] chenjianhao@pku.edu.cn and xilin@pku.edu.cn

Keywords: graphene, quantum Hall effect, quantum resistance standard

*We have studied the breakdown of the integer quantum Hall (QH) effect with fully broken symmetry, in an ultra-high mobility graphene device sandwiched between two single crystal hexagonal boron nitride substrates. The evolution and stabilities of the QH states are studied quantitatively through the nonlinear transport with dc Hall voltage bias. The mechanism of the QH breakdown in graphene and the movement of the Fermi energy with the electrical Hall field are discussed. This is the first study in which the stabilities of fully symmetry broken QH states are probed all together. Our results raise the possibility that the $\nu = \pm 6$ states might be a better target for the quantum resistance standard.*

**Introduction**

Graphene, a single layer of graphite, has continued to attract great attention from the scientific and technical communities due to the rich physics of Dirac fermions and the great potential for technological applications [1]. Started from the half-integer and the four fold degeneracy without symmetry breaking of spin or valley degrees of freedom, the quantum Hall (QH) resistance in graphene is expected to be the series of filling factor $\nu = \pm 2, \pm 6, \pm 10$……., and those Hall plateaus have been experimentally observed since 2005 [2, 3]. With the technique of suspending graphene [4] and transferring graphene onto single crystal hexagonal boron nitride (hBN) [5], one can suppress the scattering of carriers from charged impurities [6] and substrate phonons [7], resulting in substantially improved carrier mobility. Symmetry-broken integer QH states and fractional QH states were soon discovered on the improved devices [8-15].

Although most of the integer QH and fractional QH studies have been carried in GaAs/AlGaAs heterostructure for its extremely high mobility, similar studies in graphene have attracted lots of attention for a variety of reasons including the peculiar band structure of graphene and the exposure of the two dimensional electron gas (2DEG) in graphene. One interesting question has been raised that if graphene can replace GaAs to be the base of the resistance standard [16, 17]. The QH effect has been applied on the resistance unit in the international system since 1990 [18]. However, a redefinition of SI basic units (Kilogram, Ampere, Kelvin and Mole) by 2018 [19] involves the resistance standard from the QH effect. Therefore, the metrology of resistance standard, including a practical (non-SI) definition for increasing potential end users, is getting more and more important. Various works have proved the universality and reproducibility of quantum Hall resistance standard metrology in graphene with accuracy as high as part-per-billion [20-24]. Quantum resistance devices based on graphene have been realized on graphene grown on silicon carbide wafer [20-26]. More importantly, QH effect can be realized in graphene at room temperature [27], which makes graphene-based QH device a very attractive choice of a practical resistance definition for routine calibration, and a potential alternative to replace GaAs based device in the SI unit definition.

The largest measurement current to which the QH effect persists is a piece of crucial information for its application in quantum resistance standard, as a source-drain current that is too small in magnitude may limit the resolution of the electrical signal. The current induced breakdown of QH effect has been widely investigated, mostly in GaAs system, and there are multiple possible explanations [28]. Interestingly, a latest breakdown study in fractional QH effect could not be explained with the existing breakdown theories [29]. Given the special band structure of graphene and its symmetry, there is no obvious prediction for the QH breakdown behavior in graphene, and detailed experimental investigations are necessary. There have been a handful of QH breakdown studies in graphene [30-33], but a thorough stability comparison between all symmetry broken QH states is missing. Currently most of the effort for understanding the QH breakdown behavior in graphene has been focused mainly on the $\nu = \pm 2$ state.

In this work, a series of fully symmetry broken QH states were realized and we studied the stability of QH states in graphene through nonlinear measurements. Comparison between the breakdown behavior and the theoretical models are discussed. Although the $\nu = \pm 2$ QH states in graphene is commonly studied for the quantized resistance, we showed that there are alternative QH states with larger breakdown current at lower magnetic field.

**Device fabrication and low temperature measurement**
The samples consist of an hBN/Graphene/hBN van der Waals heterostructure built via a dry transfer technique adopted from reference [34, 35]. Flakes of graphene and hBN were mechanically exfoliated onto $SiO_2$(300 nm)/Si wafers. The number of layers in each graphene flake was identified by optical contrast and was later confirmed by quantum Hall measurements. The thickness of hBN flakes was measured by Atomic Force Microscopy in tapping mode. A polycarbonate/polydimethylsiloxane (PC/PDMS) bilayer was used to transfer graphene and hBN crystals; the heterostructure was assembled in inert atmosphere [34]; then the samples were fabricated via standard electron-beam lithography technique into double-gated field effect devices with Au/Cr electrodes, using hydrogen silsesquioxane (HSQ)/PMMA bilayer as etch mask [35].

The final device geometry consists of a Hall-bar shaped hBN/graphene/hBN stack placed on a highly doped silicon substrate with 300 nm $SiO_2$. The top hBN is rectangular shape, and the multiple leads of the graphene Hall-bar was extended outside of the top hBN and then covered and contacted by Au/Cr electrodes. Another Au/Cr electrode is deposited above the top hBN, acting as the top gate electrode. Doped silicon below the lower hBN and $SiO_2$ layer acts as back gate electrode (see Fig. 1(a) and 1(b)).

The measurements were performed using a standard lock-in technique at 17 Hz and with small excitation less than 13 nA. The sample was cooled in a dilution refrigerator (Leiden Cryogenics BV MNK126-450 system) with a base temperature < 6 mK and a base electron temperature < 20 mK. The temperature labeled in this work is the fridge temperature. Perpendicular magnetic fields were applied to the sample. Hall resistance $R_{xy}$ and longitudinal resistance $R_{xx}$ were measured through the four-wire configuration from Hall-bar geometry. The mobility of this device is $2 \times 10^6$ cm$^2$ V$^{-1}$ s$^{-1}$ at carrier density $1 \times 10^{12}$ cm$^{-2}$ and at sample temperature of 600 mK.

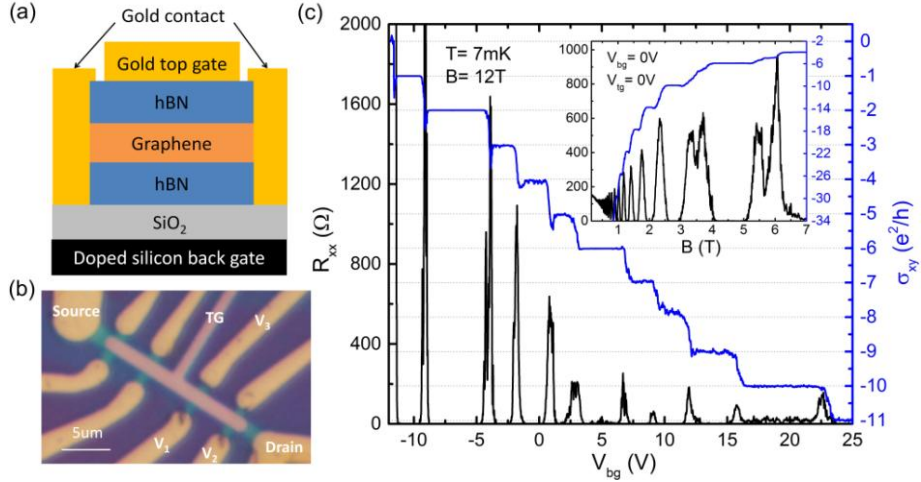

**Figure 1 (a)** Schematic diagram of the hBN/graphene/hBN sandwiched device on a doped silicon substrate. **(b)** Optical microscope image of the device, with multiple contacts in a Hall-bar geometry. The channel length $l$ (between $V_1$ and $V_2$) and width $w$ (between $V_1$ and $V_3$) are 4.675 $\mu$m and 1.57 $\mu$m, respectively. **(c)** Hall traces as a function of back gate at 12 T. The Hall conductivity is calculated from the tensor relation $\sigma_{xy} = R_{xy}/(R_{xy}^2+(w/l)^2 R_{xx}^2)$, where w and l is the channel width and length, respectively. Inset: Hall traces as a function of magnetic field. Black curves: longitudinal resistance $R_{xx}$; blue curves: Hall conductivity $\sigma_{xy}$.

**Fully symmetry broken QH effect**

Fig. 1(c) shows the longitudinal resistance $R_{xx}$ and the Hall conductivity $\sigma_{xy}$ as a function of back gate voltage $V_{bg}$ at 7 mK and 12 T. The carrier density is tuned by both the back gate and the top gate with a fixed ratio, to maintain zero perpendicular electrical field within the 2DEG. A non-zero electric displacement field is generally induced for single gate device and is theoretically expected to cause the Dirac fermions in graphene to become massive [36], likely reducing the mobility and the quantum Hall gap. The signature plateaus at filling factor ν = -2,

-6 and -10 are well defined with corresponding near-zero $R_{xx}$. The Fig. 1(c) inset panel also shows the fully developed $\sigma_{xy}$ plateaus and the distinct Shubnikov-de Hass oscillations in $R_{xx}$ at $V_{bg} = V_{tg} = 0$ V.

The $\sigma_{xy}$ plateaus and near-zero of $R_{xx}$ at other integer fillings are also developed. In Fig. 1(c), QH states from -1 to -11 are shown, which indicate that the four-fold degeneracy of spin and valley in this graphene sample has been lifted. The lifting of the four degeneracy of graphene is only seen in ultra-high quality graphene samples, and is shown to be neither spin polarized nor valley polarized in some of the emerging integer filling factors [11]. The origin of such spontaneous breaking of symmetry in quantum Hall states in single layer graphene is likely due to Coulomb interactions, but it is still under debate [11, 12]. Recently, it is proposed that some of the symmetry broken quantum Hall edge states might harbor Majorana zero modes [37], further pointing to the importance of understanding such symmetry broken QH states.

Previously, after the observation of the normal series $\nu = \pm 2, \pm 6, \pm 10$, the symmetry broken states $\nu = \pm 1$ and $\nu = \pm 3$ was quickly observed [8-12, 14, 15, 38]. However, the observation of the full integer series (e.g., $\nu = \pm 1, \pm 2, \pm 3, \pm 4 \ldots$) [11] and comparison between the stability of all the QH states was rare. As shown in Fig. 1(c), the plateau width of even filling factor state $\nu = -8$ is not necessary stronger than some odd filling factor states at 7 mK. All above indicates that it is necessary to look into the breakdown condition of the full QH sequence. For the application of graphene QH effect as the quantum resistance standard, although the $\nu = \pm 2$ states are stable and well studied, larger filling factors require much smaller magnetic field and are of great interest for practical utilization. Note that the electrical contacts to graphene in the electron-doped regime are better than that of the hole-doped regime, which could be improved with graphite contacts [39]. Our subsequent discussions are mostly based on data in the electron-doped regime.

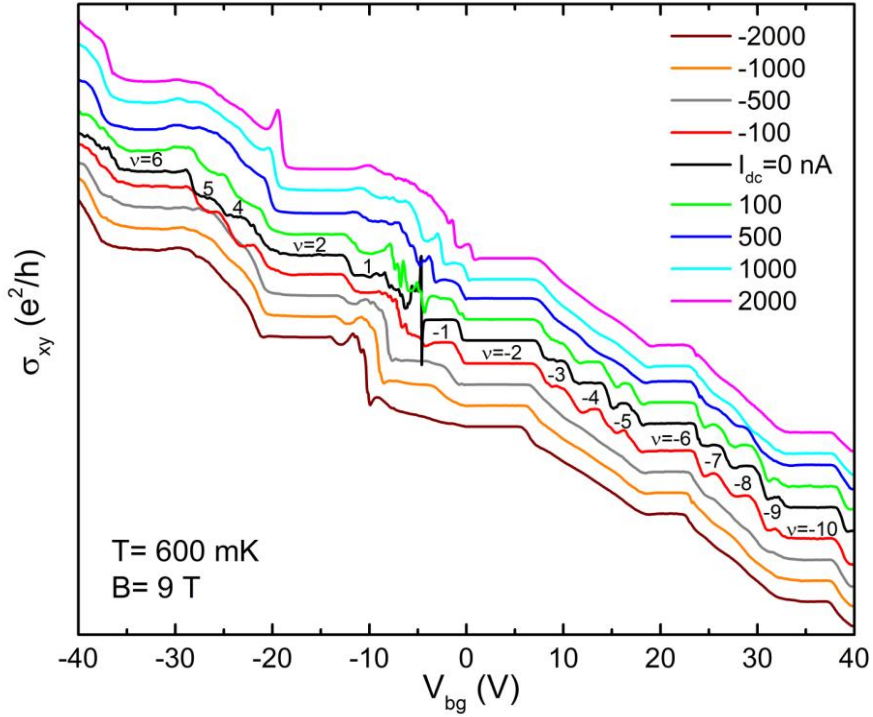

**Figure 2** Nonlinear transport measurement of graphene at 600 mK and 9 T. Quantized Hall conductivity are plotted as a function of back gate voltage at different applied dc bias currents, with an ac measurement current no larger than 13 nA. The plots are stacked with a constant offset for clarity. The midpoints of QH plateaus shift slightly as well, indicating movement of Fermi energy under applied dc bias current.

**Nonlinear transport of the QH States in graphene**

At temperature up to 600 mK, fully symmetry broken QH states can still be observed in this device, as shown in Fig. 2, and the influence of nonlinear transport is plotted. A small ac current is used for measurement and a dc bias is applied to break down the QH states. Compared with dc measurements at different excitations, ac measurement can easily guarantee that all the data have similar uncertainty. Nonlinear transport has been widely utilized in the QH breakdown research in GaAs/AlGaAs heterostructures [28] and has also been used in the depinning study of charge density wave in 2DEG [40].

In Fig. 2, the carrier density and the filling factor under certain perpendicular magnetic field are tuned by the back gate only. For the purpose of simulating the simplest quantum resistance device, our study of QH breakdown does not involve measurement with dual gates. When a dc current is added to the source-drain voltage, the stability of the QH plateaus is affected and can be observed as the shrinkage of the QH plateau width in the $\sigma_{xy}$ vs $V_g$ plot. For example, the ν = -4 and ν = -8 plateaus disappear at $I_{dc}$=500 nA.

When a dc bias is applied, several consequences can cause the breakdown of QH states. Electron heating [41], delocalization of localized state due to the Hall field [42], inter-Landau-level scattering [43], electron-phonon interaction which causes the electrons to move across the 2DEG [44] and avalanche-type breakdown of the incompressible regions [45] can all destroy the QH effect; however, determining the dominant cause in a particular device has proven to be challenging. All the above models predicts higher breakdown current with larger sample size, which agrees with the observations in integer QH effect [28] and disagrees with a recent work in fractional QH effect [29]. In short, it is difficult to predict the QH breakdown behavior in graphene based on the previous experience from GaAs/AlGaAs 2DEG, and more experiments are necessary.

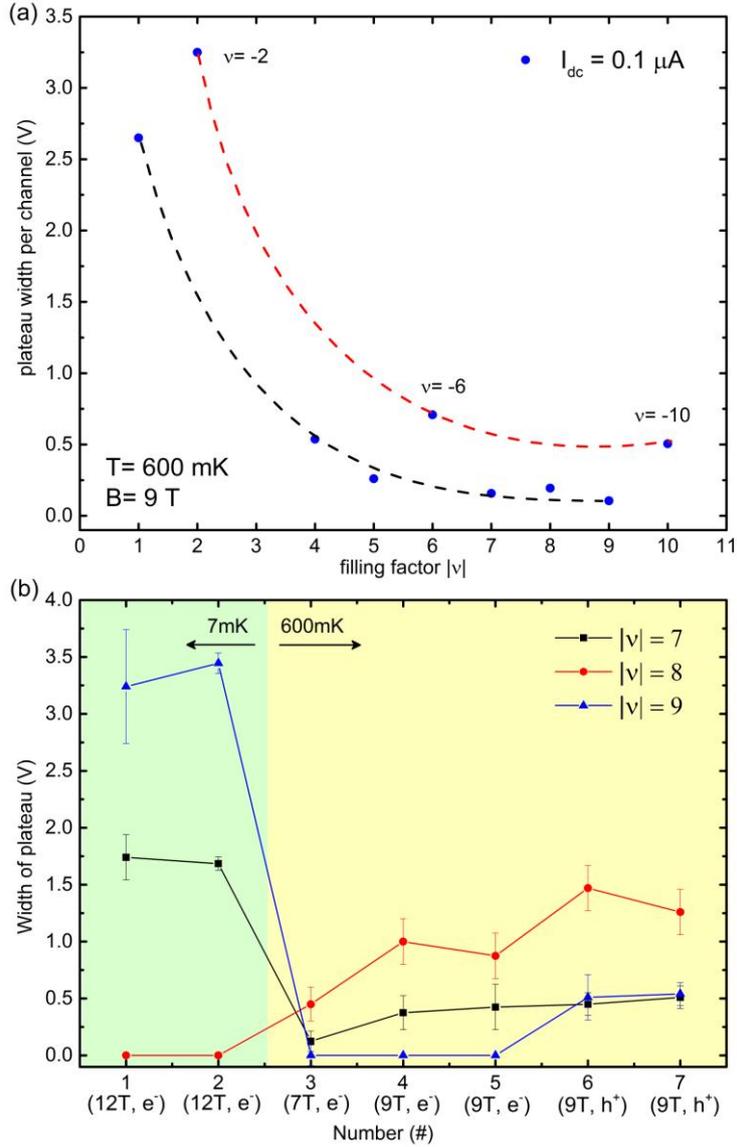

**Figure 3** The relationship between the plateau width and the filling factor **(a)** Widths of QH plateaus over filling factors are plotted as a function of filling factor at a dc bias current of 0.1 μA, analyzed from Fig. 2. Dashed lines are the guide to the eyes. **(b)** The comparison of the QH plateau widths at filling factors $|\nu|$ = 7, 8 and 9. Data #3, 4, 5 are taken by sweeping the back gate only, while the others are all taken under dual gates mode. The plateau widths of #3, 4, 5 have been divided by 2 in order to compare with the results of the dual gates mode. (12T,

e⁻) represents that the measurement was done at magnetic field 12 T on the electron side. "h⁺" represents holes.

**Analysis of QH breakdown behavior**

In Fig. 3(a), the plateau widths are normalized by the filling factor at 600 mK, so that the widths are compared at the same electron density. A higher electron density corresponds to a smaller Hall resistance, and thus a smaller heating effect from the bias current. From the clearly two trends of eye guides in Fig. 3(a), one for ν =2, 6, 10 and one for the others, the stability of different QH states depend on the Landau level's energy gap. The normal series ν = -2, -6 and -10 are stronger than other states as expected. The red dash line, indicating plateau width normalized with filling factor, decreases slower than linear as a function of filling factor and tends to saturate, which means the plateau width can increase with filling factor at certain bias current value. Indeed, the plateau width of the ν = -6 state is larger than that of the ν = -2 state at a dc bias current of 9 $\mu$A. For a given magnetic field, the higher the Landau level corresponds to the higher electron density. At 600 mK, the hysteresis from gate sweeping only changes the position of the plateau in $V_{bg}$ but not the plateau width. Therefore, the enlargement of the width at high normal filling factor is from the electron density effect but not from the nonlinear gate response.

The ν = -3 is the weakest state and cannot survive even at the bias current as small as 0.1 $\mu$A at 600mK, which may relates to a transition to a new broken symmetry phase [9] caused by the Hall electric field from bias voltage. Kosterlitz-Thouless transition has been expected in the both ν = ±3 and ν = ±5 state, but the stability of these states in our device are very different [46]. The fact that ν = -8 state is even weaker than ν = -3 state at 7 mK is unexpected (Fig. 1(c)). In addition, ν = -8 state is weaker than ν = -7 and ν = -9 states is also unexpected. The ν = ±8 state only breaks spin or valley symmetry, so it is supposed to be easier to develop than the ν = ±7 or ν = ±9 states. Fig. 3(b) shows the width of the ν = ±7, ±8, ±9 plateaus from different sets of measurements, at different magnetic fields and at two different temperatures (7 mK and 600 mK). It's clear that the ν = ±8 state is always stronger than the ν = ±7 and ν = ±9 states for all the conditions we have at 600 mK, contrary to the situation at 7 mK, where the ν = -8 state is the weakest. The anomaly of the ν = -8 state at ultra-low temperature cannot be understood with the mechanism of QH breakdown. Such anomaly implies a possible scenario: a novel phase such as bosonic excitation competes with the conventional integer QH ν = -8 state in graphene. Such a novel phase is recently suggested at the zero temperature limit with Majorana zero modes in the ν = 8 QH state without involving superconductivity [37].

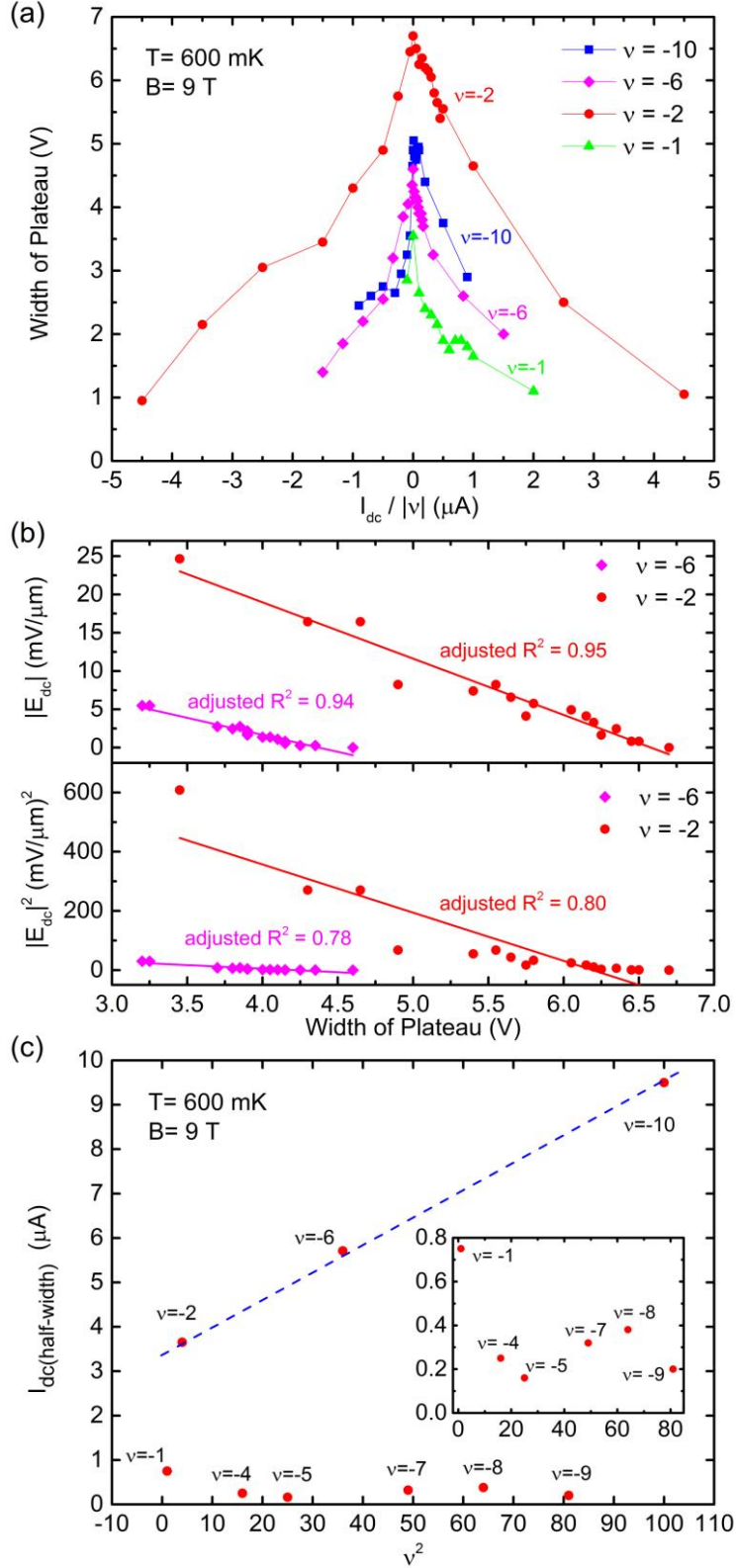

**Figure 4** The relationship between the electrical field, breakdown current and the filling factor, analyzed from Fig. 2. **(a)** Widths of QH plateaus are plotted as a function of dc bias current over filling factors ($I_{dc}/|\nu|$ for $\nu$ = -1, -2, -6 and -10). **(b)** Upper panel is $|E_{dc}|$ as a function of plateau width, while lower panel is $|E_{dc}|^2$ as a function of plateau width. $|E_{dc}|=|I_{dc}|*R_{xy}/w$ is the transverse electric field, introduced by dc bias current, across the sample.
8

Linear fits of both ν = -2 and -6 and the corresponding adjusted R-square are shown in the figure, where the linear relation in the upper panel fits our data better than that of the lower panel. The adjusted R-square is a modified version of R-square, which assesses the goodness of fits better than the latter. **(c)** The breakdown current at half-widths (the bias current when a QH plateau width decreases to 50% of its largest width) are plotted as a function of $\nu^2$ from ν = -1 to -10 QH states (except for ν = -3, which is too weak to do the same analysis). The dashed blue line is guide to the eyes. Inset: the zoom-in for the low $I_{dc(half\text{-}width)}$ data. There is no definitive relation for the ν = -1, -4, -5, -7, -8 and -9 states.

With the dc bias current, the QH plateaus are destabilized. As shown in Fig. 4(a), the x-axis is bias current normalized by the filling factor, in order to compare the effect of the electrical field. The electrical field across the Hall-bar is equal to the bias voltage over the width, which is the product of bias current $I_{dc}$ and the Hall resistance $h/(|\nu|e^2)$. The plateau widths in Fig. 4(a) are almost linear to the electrical field with almost identical slope. As further elaborated in Fig 4(b), we can determine that the plateau width strongly depends on the linear term of the electrical field across the sample, rather than the square term at the zero field limit. If the breakdown of QH states in graphene associates with the electrical field across the sample, then the breakdown mechanism may be related to the coherent many-electron inter-Landau-level scattering [43], where a critical Hall field linear term multiplying with the electron charge plays an important role by comparing with the Landau level energy.

We tested all the contacts and pick up the configuration with the lowest noise for this measurement. Therefore, the breakdown current provided from this work is only from a given pair of contacts. If the breakdown dominantly originates from the heating effect, the power from the heating should scales with $\sigma_{xx}(E_x^2 + E_y^2) \approx \sigma_{xx}E_y^2$, which is contracted to Fig. 4(b).

The mechanism of the electro-phonon interaction can also be excluded because it happens when the electron drift velocity $E_y/B$ exceeds the sound velocity in graphene, which should cause similar breakdown electrical field for all QH states at fixed magnetic field. The model of delocalization of localized state due to the Hall field predicts the electrical field scales with $B^{1.5}$, which is not observed in our measurement either. The avalanche-type breakdown involves metastable states with different number of metallic paths, and hysteresis is expected, but hysteresis is not clearly observed as well at 600 mK in our nonlinear transport. There is some small hysteresis from gate sweeping at the lowest temperature, which can be attributed to movement, charging and discharging of the charge trap in the silicon substrate.

Finally, the half-widths of the breakdown current are plotted as a function of $\nu^2$ in Fig. 4(c) and there is a strong linear dependence among the normal series ν = -2, -6 and -10. The breakdown currents where the plateaus of ν = -6 and ν = -10 shrink to their half-widths are larger than that of the ν = -2 state. This observation suggests that although the ν = -2 state's plateau is stronger than the others, its persistence at high current bias is not necessary the best. Therefore, the breakdown current directly measured at the center of the QH plateau are presented and discussed in the next section.

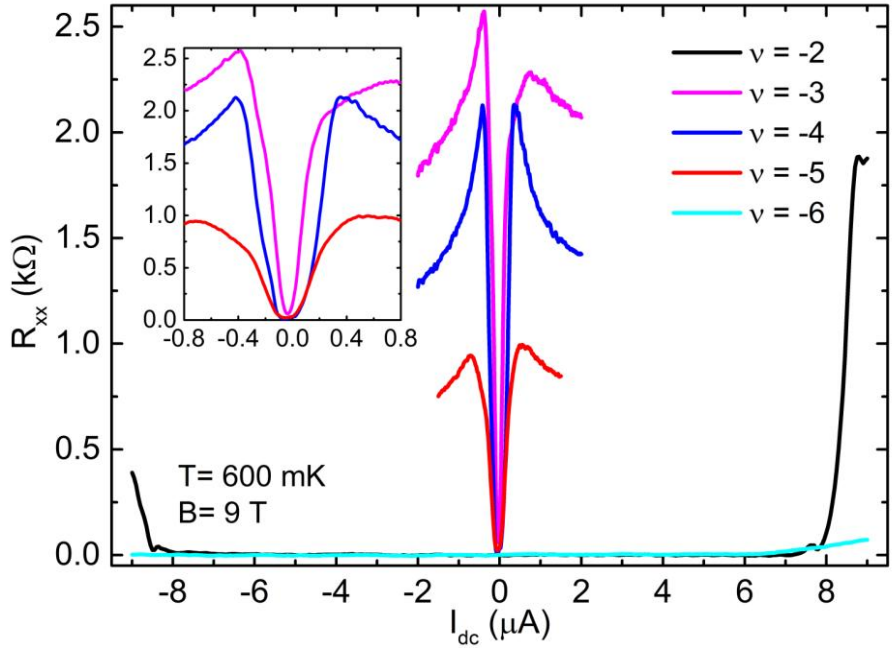

**Figure 5** The longitudinal resistance as a function of bias current for different QH states. The gate voltages are fixed at the center of the plateaus from $\nu = -2$ to $\nu = -6$. Inset is the zoom-in plot for the weaker plateaus.

**Quantum resistance standard based on the $\nu = \pm 6$ state**

In Fig. 2, different QH plateaus reduce their widths differently with bias voltage. In order to directly evaluate the breakdown current quantitatively at different QH states, Fig. 5 is provided. The states without lifting the spin or valley degeneracy are much stronger, as expected. However, in term of the breakdown current, there is no significant difference between the $\nu = -6$ state and the $\nu = -2$ state.

In GaAs/AlGaAs heterostructure, the $\nu = 2$ state usually has the strongest energy. The energy gap of the $\nu = 1$ state results from the Zeeman energy, which is much smaller than the Landau energy in GaAs 2DEG; the Landau levels higher than 2 will be less stable. In graphene, $\nu = \pm 2$ state is the first state of the regular series and is well studied either for the quantum resistance standard or the QH stability [20-24, 30-33, 47]. If the breakdown current in graphene has no significant difference between $\nu = -6$ and the $\nu = -2$, then the resistance metrology based on the $\nu = \pm 6$ state has its advantage on the application. For a given electron density, the $\nu = \pm 6$ state requires only 1/3 of the magnetic field of the $\nu = \pm 2$ state, so $\nu = \pm 6$ may be a better state than $\nu = \pm 2$ for practical resistance definition of routine calibration. However, it is worth noting that lower magnetic field results in worse quantization condition and lower resistance, which requires higher sensitivity of the voltage probes; thus, further experiments are needed for a comprehensive comparison between the accuracy of the $\nu = \pm 6$ and the $\nu = \pm 2$ states in graphene as the new resistance standard.

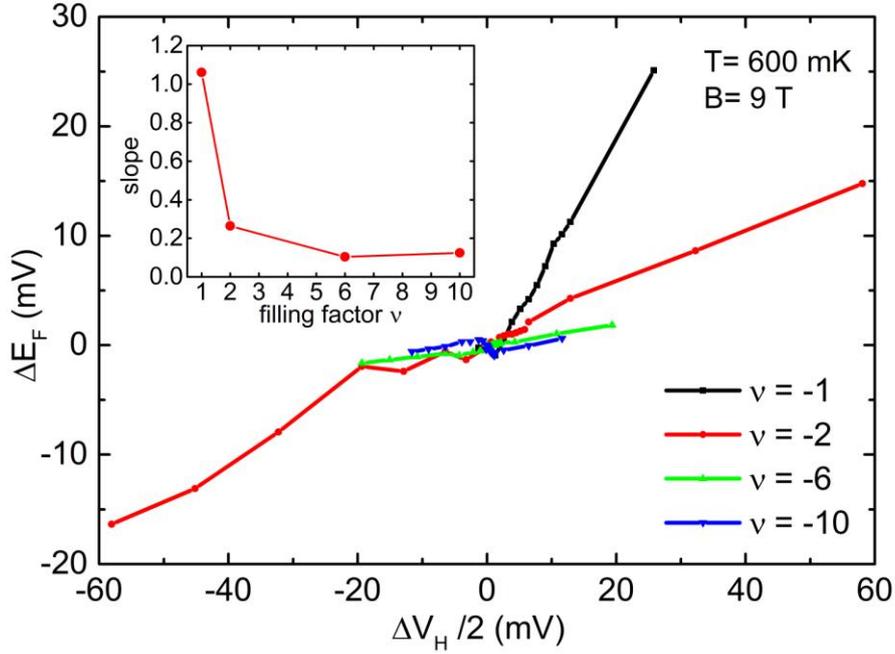

**Figure 6** Shifts of the Fermi energies of QH plateaus versus half of the Hall voltage applied. The Fermi energies of QH plateaus are defined as the middle points of QH plateaus in the $R_{xx}(V_g)$ curve; the different filling factors are colored as black ($\nu = -1$), red ($\nu = -2$), green ($\nu = -6$) and blue ($\nu = -10$). The Hall voltage can be obtained from $\Delta V_H = I_{dc} V_{H,ac}/I_{ac}$. The Fermi energy shift vs Hall voltage curve of the $\nu = -10$ plateau initially has a negative slope and then the slope becomes positive. The mechanism for such change needs further investigation. Inset: Slopes of the $d\Delta E_F/d(\Delta V_H/2)$ curves for filling factors $\nu = -1, -2, -6$ and $-10$.

**Movement of the Fermi energy as a function of Hall field**

In addition to the breakdown of the QH states, the application of a large dc bias (or a large Hall voltage, equivalently) resulted in a shift of the Fermi energy of each Landau level, which can be measured experimentally as the shift of the gate voltage at the middle point of a quantum Hall plateau. Fig. 6 shows the dependence of the shift of the Fermi energy ($\Delta E_F$) on one half of the Hall voltage ($\Delta V_H/2$) in the device for filling factor $\nu = -1, -2, -6$ and $-10$. It can be seen that $\Delta E_F$ depends almost linearly on $\Delta V_H/2$ for $\nu = -1, -2, -6$, and the slopes tend to decrease with increasing filling factors (Inset of Fig. 5). Such shifts of Fermi energies are only observed for filling factor $\nu = -1, -2, -6$ and $-10$. Other QH states are destroyed before the dc bias is large enough to noticeably shift the Fermi energies (see Fig. 2). This is the first time that such shifts of Fermi energies under large dc bias have been reported in two-dimensional QH systems. One possible reason for the shift is that a large Hall voltage effectively tilts the energy landscape of the sample in the transverse direction, and because of the small density of state in graphene as compares to other 2DEG systems, such tilts in energy landscape result in an effective doping of the sample. Under this assumption, $\Delta E_F = k*\Delta V_H/2$ is expected, with the slope $k = 1$ for all the filling factors. However, experimentally we only observed $k \sim 1$ at $\nu = -1$, and $k$ for other filling factors is smaller than one and different (Inset of Fig. 6). Furthermore, the $k$ value for the $\nu = -10$ plateau is initially negative and then becomes positive. The physical origin of such filling factor dependent shift of the Fermi energy as a function of dc bias needs to be pursued by further experimental and theoretical investigations.

**Conclusion**

In summary, we have observed a series of fully symmetry broken QH states in our hBN/Graphene/hBN device and their stability were studied through nonlinear transport. By comparing the breakdown behaviors of the normal series $\nu = \pm2, \pm6, \pm10$, we suggest it is worthwhile to consider the possibility of using the $\nu = \pm6$ states in graphene as quantum resistance standard.

**Acknowledgement:**

We are grateful to Haiwen Liu, Hua Jiang, Fa Wang, Xiangang Wan and X. C. Xie for their helpful comments. This project has been supported by the National Basic Research Program of China (NBRPC Grant Nos. 2013CB921900, 2014CB920900, 2015CB921101) and the National Natural Science Foundation of China (NSFC Grant Nos. 11274020, 11374021, 11322435). K.W. and T.T. acknowledge support from the Elemental Strategy Initiative conducted by the MEXT, Japan and a Grant-in-Aid for Scientific Research on Innovative Areas "Science of Atomic Layers" from JSPS.